\documentclass[aps,showpacs,twocolumn,showkeys,prb]{revtex4}
\usepackage{graphicx,color,epsfig,rotate}
\usepackage{times,xspace}
\usepackage{amsbsy,amssymb,amsmath,bm}
\usepackage{dcolumn}
\usepackage{bm}
\input{epsf.sty}

\def\bbbc{{\mathchoice {\setbox0=\hbox{$\displaystyle\rm C$}\hbox{\hbox
to0pt{\kern0.4\wd0\vrule height0.9\ht0\hss}\box0}}
{\setbox0=\hbox{$\textstyle\rm C$}\hbox{\hbox
to0pt{\kern0.4\wd0\vrule height0.9\ht0\hss}\box0}}
{\setbox0=\hbox{$\scriptstyle\rm C$}\hbox{\hbox
to0pt{\kern0.4\wd0\vrule height0.9\ht0\hss}\box0}}
{\setbox0=\hbox{$\scriptscriptstyle\rm C$}\hbox{\hbox
to0pt{\kern0.4\wd0\vrule height0.9\ht0\hss}\box0}}}}

\begin{document}
\title{Using magnetostriction to measure the spin-spin correlation function and magnetoelastic coupling in the quantum magnet NiCl$_2$-4SC(NH$_2$)$_2$}

\author{V. S. Zapf$^1$, V. F. Correa,$^{2,{\dagger}}$ P. Sengupta,$^{1,3}$ C. D. Batista,$^3$ M. Tsukamoto,$^4$, N. Kawashima,$^4$ P. Egan,$^5$, C. Pantea,$^1$ A. Migliori,$^1$ J. B. Betts,$^1$ M. Jaime,$^1$ A. Paduan-Filho$^6$}

\affiliation{$^1$National High Magnetic Field Laboratory (NHMFL), Los Alamos National Lab (LANL), Los Alamos, NM \\
$^2$NHMFL, Tallahassee, Florida \\
$^3$Condensed Matter and Thermal Physics, LANL, Los Alamos, NM\\
$^4$ Institute for Solid State Physics, University of Tokyo, Kashiwa, Chiba, Japan \\
$^5$ Oklahoma State University, Stillwater, OK\\
$^6$ Instituto de Fisica, Universidade de Sao Paulo, Brazil \\
$^{\dagger}$ Now at Comisi\'on Nacional de Energ\'ia At\'omica,
Centro At\'omico Bariloche, 8400 S. C. de Bariloche, Argentina}

\date{\today}

\begin{abstract}

We report a method for determining the spatial dependence of the
magnetic exchange coupling, $dJ/dr$, from magnetostriction
measurements of a quantum magnet. The organic Ni $S = 1$ system
NiCl$_2$-4SC(NH$_2$)$_2$ exhibits lattice distortions in response
to field-induced canted antiferromagnetism between $H_{c1} = 2.1$
T and $H_{c2} = 12.6$ T. We are able to model the magnetostriction
in terms of uniaxial stress on the sample created by magnetic
interactions between neighboring Ni atoms along the c-axis. The
uniaxial strain is equal to $(1/E)dJ_c/dx_c \langle S_{\bf r}
\cdot S_{{\bf r}+ {\bf e}_c} \rangle$, where $E$, $J_c$, $x_c$ and
${\bf e}_c$ are the Young's modulus, the nearest neighbor (NN)
exchange coupling, the variable lattice parameter, and the
relative vector between NN sites along the c-axis. We present
magnetostriction data taken at 25 mK together with Quantum Monte
Carlo calculations of the NN spin-spin correlation function that
are in excellent agreement with each other. We have also measured
Young's modulus using resonant ultrasound, and we can thus extract
$dJ_c/dx_c = 2.5$ K/$\AA$, yielding a total change in $J_c$
between $H_{c1}$ and $H_{c2}$ of 5.5 mK or 0.25\% in response to
an 0.022\% change in length of the sample.

\end{abstract}

\maketitle

In many insulating magnets, the magnetic coupling is caused by
superexchange interactions created when atomic or molecular
orbitals overlap. Since the radial dependence of the orbital wave
functions can be quite steep, the overlap integrals and the
resulting exchange coupling $J$ depend strongly on the interatomic
bond lengths $r$. Past experiments have probed dependence of $J$
on $r$ using hydrostatic pressure or chemical substitution to vary
the bond length, and Raman spectroscopy to measure $J$. These
results were combined with high-intensity X-ray scattering
measurements or elastic neutron scattering to determine the bond
lengths. \cite{Aronson91,Massey90,Harrison80,Cooper90}

\begin{figure}[tbp]
\epsfbox{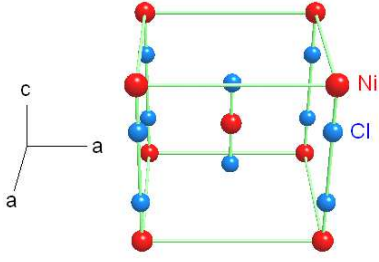}
\caption{Unit cell of tetragonal
NiCl$_2$-4SC(NH$_2$)$_2$ showing Ni (red) and Cl (blue) atoms. The
thiourea molecules have been omitted for clarity.}
\label{crystalstructure}
\end{figure}

Here we demonstrate a simple and novel approach to investigating
the spatial dependence of the superexchange interaction in the
quantum magnet NiCl$_2$-4SC(NH$_2$)$_2$ (DTN). We use applied
magnetic fields to create an effective pressure and measure the
response of the soft organic lattice via magnetostriction. The $S
= 1$ Ni ions in DTN form a body-centered tetragonal structure
\cite{LopezCastro63} shown in Fig. \ref{crystalstructure}. The
dominant magnetic superexchange interaction $J_c = 2.2$ K is
antiferromagnetic (AFM) and occurs along linear Ni-Cl-Cl-Ni bonds
in the tetragonal c-axis.\cite{Zapf06,Zvyagin07} Along the a-axis,
$J_a = 0.18$ K is an order of magnitude smaller and no diagonal or
next-nearest neighbor couplings have been found within the
resolution of inelastic neutron scattering measurements.
\cite{Zapf06} We thus treat this compound as a 1D system of
Ni-Cl-Cl-Ni chains only weakly coupled in the a-b plane. Because
$J_c$ is sensitive to the Ni inter--ion bond lengths, a magnetic
stress is created between adjacent Ni spins along the c-axis. This
stress depends on the relative orientation of the two spins, e.g.
on the NN spin--spin correlation function $\langle S_{\bf r} \cdot
S_{{\bf r}+ {\bf e}_c} \rangle$.

\begin{figure}[tbp]
\centering
\epsfbox{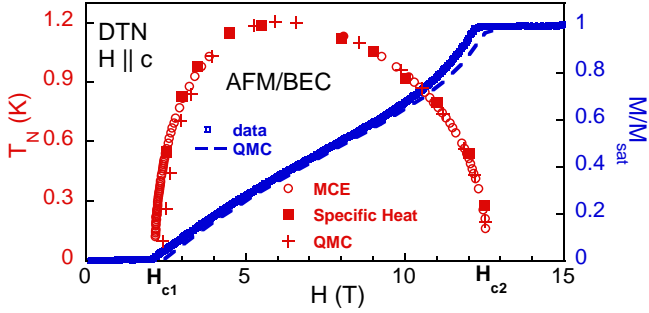}
\caption{Temperature $T$ -
Magnetic field $H$ phase diagram for $H || c$ determined from
specific heat and magnetocaloric effect (MCE) data, together with
the result of Quantum Monte Carlo (QMC) simulations.
\cite{Zapf06,Zvyagin07} The magnetization vs field measured at 16
mK and calculated from QMC is overlayed onto the phase diagram.
\cite{PaduanFilho04}} \label{introduction_phasediagram}
\end{figure}

In DTN, the NN spin--spin correlation function varies strongly with
magnetic field. DTN exhibits AFM order for applied
fields along the c--axis between $H_{c1} = 2.1$ T and $H_{c2} =
12.6$ T and with a maximum N\'{e}el temperature of $T_N = 1.2$ K,
as shown in the phase diagram in Fig.
\ref{introduction_phasediagram}. The AFM order is
confined to the a-b plane at $H_{c1}$. However, as the field is
increased from $H_{c1}$ to $H_{c2}$, the spins cant along the
c-axis and finally saturate for $H > H_{c2}$. This is illustrated
in the magnetization vs field curve shown in Fig.
\ref{introduction_phasediagram}.

The lack of magnetic order at zero field is due to a strong
easy--plane uniaxial anisotropy that creates a splitting $D$ at
zero field between the $S_z = 0$ ground state and the $S_z = \pm
1$ excited states of the Ni ion. In applied fields parallel to the
tetragonal c-axis, the Zeeman effect then lowers the $S_z = -1$
state until it becomes degenerate with the $S_z = 0$ state,
resulting in a magnetic ground state and AFM order below the
N\'{e}el temperature. \cite{Zvyagin07} Since the $S_z = -1$ state
is broadened by AFM dispersion, the region of overlap between $S_z
= -1$ and $S_z = 0$ extends from $H_{c1} = 2.1$ T up to $H_{c2} =
12.6 $ T.

Here we show that the bare NN spin-spin correlation function can
be directly proportional to the c-axis magnetostriction. Since all
the terms in the magnetic Hamiltonian of this compound have been
measured, we can calculate $\langle S_{\bf r} \cdot S_{{\bf r}+
{\bf e}_c} \rangle$ using Quantum Monte Carlo simulations to
predict the magnetostriction response as a function of the applied
magnetic field. By combining these results with resonant
ultrasound spectroscopy to determine the elastic moduli, we are
also able to extract the leading linear term in the spatial
dependence of the exchange interaction $J_c(r)$ along the
tetragonal c-axis.

\epsfxsize=150pt

\epsfxsize=250pt

\begin{figure}[tbp]
\epsfbox{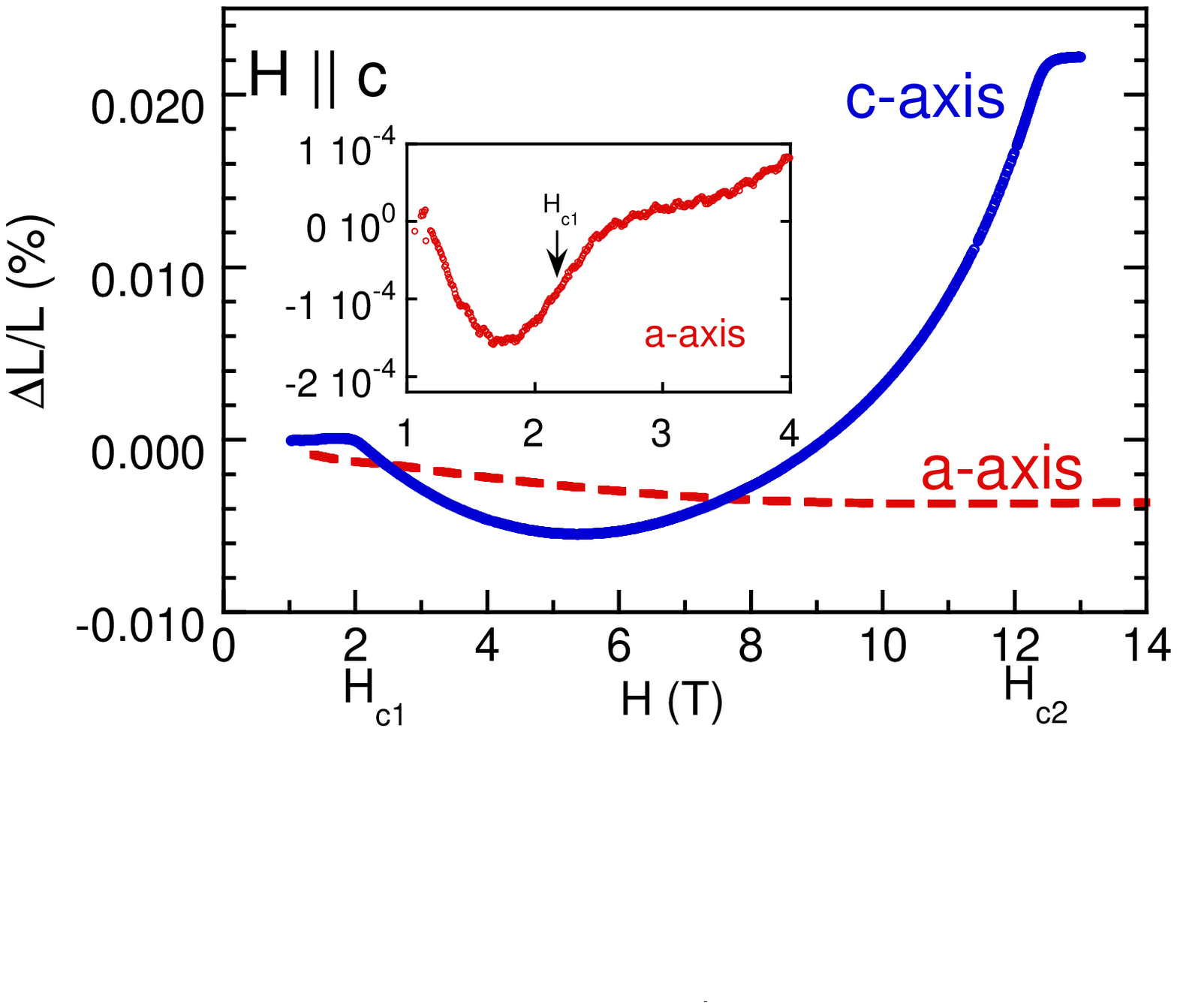}
\caption{Normalized percentage
length change \%$\Delta L/L$ as a function of magnetic field
measured along the crystallographic c-axis (solid blue lines) and
a-axis (dashed red lines). The data is taken at $T = 25$ mK with
the magnetic field applied along the c-axis. The inset shows the
feature at $H_{c1}$ in \%$\Delta L_a/L_a$ in greater detail, and a
straight line has been subtracted from the inset data for clarity.}
\label{Magnetostriction}
\end{figure}

We first present magnetostriction measurements that were performed
on single crystals of DTN down to 25 mK in a 20 T magnet at the
National High Magnetic Field Laboratory in Tallahassee, FL, as
described in Ref. 9. The magnetostriction as a function of $H$ for
$H || c$ is shown in Fig. \ref{Magnetostriction} for both the a
and c-axes of the crystal. The c-axis magnetostriction $\Delta
L_c/L_c$ shows sharp shoulders at the boundaries of the ordered
state at $H_{c1}$ and $H_{c2}$, and nonmonotonic behavior in
between. The net difference between the c-axis lattice parameter
at $H_{c1}$ and $H_{c2}$ is $0.022$\%. The nonmonotonic behavior
of the magnetostriction contrasts with the roughly linear
dependence of the magnetization $M(H)$ in the region of AFM order
between $H_{c1}$ and $H_{c2}$, as shown in Fig.
\ref{introduction_phasediagram}. It also contrasts with the
magnetostriction observed in the Cu dimer spin gap system
KCuCl$_3$, in which the magnetostriction closely tracks the
magnetization. \cite{Sawai05}

The a-axis lattice parameter decreases monotonically by an amount
that is an order of magnitude smaller than the change in the
c-axis parameter, reflecting the fact that $J_a << J_c$. The
a-axis behavior is more difficult to explain since the exchange
interaction is mediated by an unknown and likely convoluted path,
and because the a-axis is subject to significant Poisson forces
from the larger c-axis distortion.

We thus focus on the c-axis magnetostriction and we suggest a
straightforward explanation for its nonmonotonic field-dependence
between $H_{c1}$ and $H_{c2}$. The canted AFM order results in two
competing forces on the c-axis of the lattice. Near $H_{c1}$, the
Ni spins order antiferromagnetically, thus creating an attractive
force. By reducing the c-axis lattice parameter, the system can
increase $J_c$, and thereby lower the energy of the
antiferromagnetically aligned spins. However, with increasing
field the spins cant, resulting in a ferromagnetic component to
the order. The ferromagnetic component stretches the lattice,
thereby reducing $J_c$. Once the magnetic field exceeds $\sim 5.5$
T, the ferromagnetic component wins and the lattice expands.

\epsfxsize=220pt

We now test this model by calculating the expected c-axis
magnetostriction. The energy density of the system can be written
as the sum of magnetic and lattice components, $\epsilon =
\epsilon_e + \epsilon_m$, with
\begin{eqnarray}
\epsilon_e &=& \frac{1}{2} E \Big ( \frac{x_c-c_o}{c} \Big )^2
\nonumber \\
\epsilon_m &=& \frac{2}{N a^2 c} \langle {\cal H}_m \rangle
\end{eqnarray}
Here $a$ and $c$ are the lattice parameters at zero field. $E$ is
Young's modulus along the $c$-axis and $N$ is the total number of
Ni sites. The factor of 2 is required because there are two Ni
atoms per unit cell of volume $a^2c$. The variable $c_o$ is the
value of the lattice parameter along the $c$-axis {\it in the
absence of magnetic interactions and external pressure}. The
variable $x_c$ is the new value of the lattice parameter when the
magnetic interactions are included. We have neglected the effect
of a-axis strain inducing changes in the c-axis via the Poisson
ratio since it is a 1 \% effect.

The magnetic Hamiltonian ${\cal H}_m$ is:
\begin{equation}
{\cal H}_m=\sum_{{\bf r},\nu}
J_{\nu}\textbf{S}_{{\bf r}}\cdot
\textbf{S}_{{\bf r}+e_{\nu}}+ \sum_{\textbf{r}} [D
(S^z_\textbf{r})^2 - g\mu_B H  S^z_\textbf{r}],
\end{equation}
where $e_\nu=\{a{\hat{\bf x}},b{\hat{\bf y}},c{\hat{\bf z}}\}$ are the relative vectors
between NN Ni ions along the a, b and c--axis respectively. In this Hamiltonian,
the magnitude of the various parameters $D$, $J_a$, $J_c$, and $g$ have been measured
experimentally via ESR and neutron diffraction in combination with
Quantum Monte Carlo simulations. \cite{Zapf06,Zvyagin07}

We can now obtain the value of $x_c$ as a function of magnetic field
by minimizing the total energy with respect to $x_c$:
\begin{equation}
\partial_{x_c} \epsilon = \frac{E}{c^2} (x_c-x_o) + \partial_{x_c}
\epsilon_m=0.
\end{equation}
We assume that the only term in ${\cal H}_m$ that depends on $x_c$
is the AFM Heisenberg coupling along the $c$--axis. The
single--ion anisotropy $D$ typically has a much smaller dependence
on $x_c$. In addition, since the temperature at which the
magnetostriction measurements were performed (25 mK) is much lower
that any characteristic energy of the system, we will assume that
$T = 0$ K. Under these conditions we obtain:
\begin{equation}
\partial_{x_c} \epsilon_m = \frac{2}{a^2c} \partial_{x_c} J|_{x_c=c}
\langle {\bf S}_{{\bf r}} \cdot {\bf S}_{{\bf r}+{\bf e}_{c}} \rangle.
\label{emin}
\end{equation}
In Eq.~\ref{emin} we have applied the
Hellman--Feynman and assumed that $\partial_{x_c} J \simeq
\partial_{x_c} J|_{x_c=c}$ because the relative distortion is very
small. Substituting into Eq. (3) we find that:
\begin{equation}
\frac{E}{c^2} (x_c-x_o) + \frac{1}{a^2c} \partial_{x_c} J|_{x_c=c}
\langle {\bf S}_{{\bf r}} \cdot {\bf S}_{{\bf r}+{\bf e}_{c}}
\rangle=0. \label{cond1}
\end{equation}
We know that $x_c = c$ for $H = 0$ and thus,
\begin{equation}
\frac{E}{c^2} (c-x_o) + \frac{1}{a^2c} \partial_{x_c} J|_{x_c=c}
\langle {\bf S}_{{\bf r}} \cdot {\bf S}_{{\bf r}+{\bf e}_{c}} \rangle_{H=0}=0,
\label{cond2}
\end{equation}
where $\langle {\bf S}_{{\bf r}} \cdot {\bf S}_{{\bf r}+{\bf e}_{c}} \rangle_{H=0}$ indicates that
the mean value is computed for a field $H=0$. By taking the difference between Eqs.~\ref{cond1} and \ref{cond2}
we obtain:
\begin{equation}
\frac{\Delta L}{L}= - \frac{\partial_{x_c}
J|_{x_c=c}}{a^2E}  [\langle {\bf S}_{{\bf r}} \cdot {\bf S}_{{\bf r}+{\bf e}_{c}} \rangle_{H=0}
- \langle {\bf S}_{{\bf r}} \cdot {\bf S}_{{\bf r}+{\bf e}_{c}} \rangle_{H}],
\label{fineq}
\end{equation}
where ${\Delta L}/L=(x_c-c)/c$. Thus our measured c-axis
magnetostriction is proportional to the NN spin-spin correlation
function with a proportionality constant of
\begin{equation}
\kappa=\frac{1}{a^2E} \partial_{x_c} J|_{x_c=c} \label{kappa}.
\end{equation}

\begin{figure}[tbp]
\epsfbox{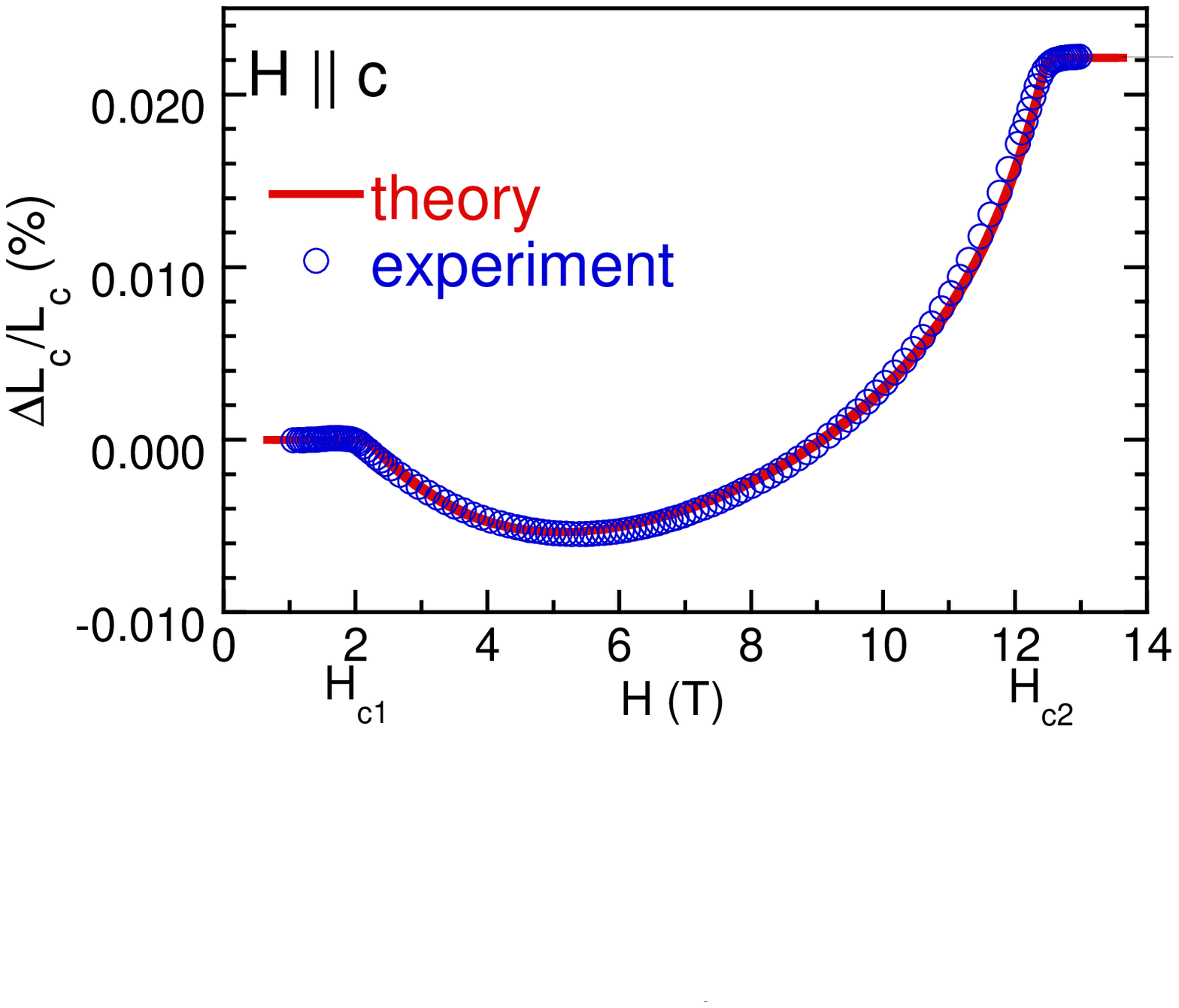}
\caption{Comparison of experimental c-axis
magnetostriction data as a function of $H$ for $H || c$ with the
model described in the text.}
\label{model}
\end{figure}

We can therefore model the experimental magnetostriction data with
the parameter $\kappa$ as the only fitting parameter. We have
determined the NN spin-spin correlation function using Quantum
Monte Carlo simulations on a $8\times8\times24$ lattice and the
parameters: $J_c = 2.2$ K, $J_a = 0.18$ K, and $D = 8.6$
K.\cite{Zvyagin07} The results of our model are shown in
comparison with the measured magnetostriction in Fig.~\ref{model},
with $\kappa= 1.00 \times 10^{-5}$. The agreement between theory
and experiment is very good and confirms our hypothesis that the
spatial dependence of $D$ is much smaller than the spatial
dependence of $J$ and can thus be neglected. This is to be
expected since $J$ results from the overlap integral between
adjacent molecular wave functions, which can have large radial
dependencies with high power-laws, whereas $D$ depends on
crystalline electric fields that change more weakly with lattice
distortions. For instance, previous experimental and theoretical
works on other compounds have modelled the spatial dependence of
superexchange interactions as a power-law $J(r) = br^{-n}$ where
$r$ is the relevant spacing between magnetic ions. Values for the
exponent $n$ of 10-14 have been reported for metal
halides,\cite{Massey90,Harrison80} and 2-7 for cuprates.
\cite{Aronson91,Cooper90} In this work, we are determining the
leading linear term in the expansion of $J(r)$, e.g. $\partial_r J
\approx -n J/r$. We have neglected higher order terms because the
relative change of the lattice parameter $c$ that results from the
magnetic stress is always smaller than 0.03\% as shown in
Fig.\ref{Magnetostriction}.

\epsfxsize=220pt

We take our analysis one step further and quantitatively determine
the spatial dependence of the AFM exchange interaction $dJ_c/dx_c$
from Eq. \ref{kappa}. The lattice parameters $a = 9.558$ \AA{} and
$c = 8.981$ \AA{} are known from published X-ray diffraction
measurements at 110 K. \cite{LopezCastro63} That leaves Young's
modulus $E$ as the remaining quantity to be determined before we
can extract $dJ_c/dx_c$.

\epsfxsize=200pt
\begin{figure}[tbp]
\epsfbox{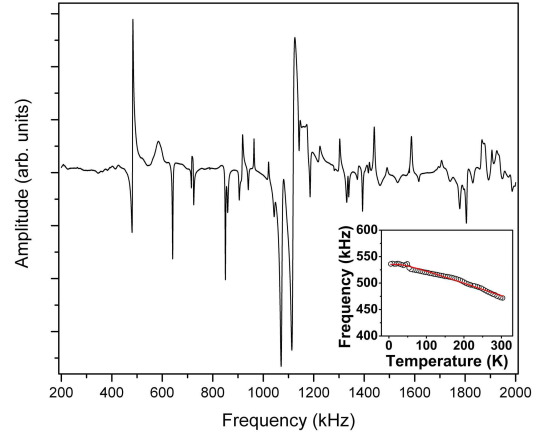}
\caption{Mechanical resonances of DTN at room temperature. Inset:
Temperature-dependence of the major peak near 500 kHz between 5 K
and 300 K. The line is a fit to the usual Einstein oscillator
model equation \cite{Varshni70} and used to extrapolate the
resonance to 0 K.}\label{RUS}
\end{figure}

Thus, we have also measured Young's modulus using Resonant
Ultrasound Spectroscopy (RUS) between 300 K and 5
K.\cite{Migliori97, Migliori93} Mechanical resonances of a roughly
cube-shaped single crystal of DTN were determined at zero field in
a He-cooled Oxford Instruments flow-cryostat. The six independent
elastic moduli were determined between 300 K and 200 K and their
values at room temperature are shown in Table
\ref{elastic_constants}.

For lower temperatures, determination of all of the resonances
used in the fitting procedure became ambiguous. However, two good
resonances could be identified down to 5 K and based on the
temperature dependencies of these resonances, we extrapolated the
value of Young's Modulus to 0 K. Young's modulus $E$ in a
tetragonal crystal is given by:
\begin{equation}
E_{33} = C_{33} - \frac{2C^2_{13}}{C_{11}+C_{12}}, \label{eq9}
\end{equation}
yielding $E = 7.5 \pm 0.7$ GPa at 0 K.

Now we can use equation (9) to calculate $\partial_x J|_{x=x_o} =
dJ_c/dx_c = 2.5$ K/$\AA$, yielding a total change in $J_c$ between
$H_{c1}$ and $H_{c2}$ of 5.5 mK or 0.25\%. This in turn results in
a 0.1\% shift in $H_{c2}$ relative to its value in the absence of
magnetostriction effects. The dominant uncertainty in these
calculations comes from the 10\% error bar in estimating Young's
modulus $E$ due to the softness of the crystal.

Previous papers about DTN have assumed that $J_c$ is constant when
calculating the critical fields, the magnetization, and other
field-dependent measurable quantities. \cite{Zapf06,Zvyagin07}
Since $J_c$ only varies by 0.25\%, these assumptions are quite
reasonable and well within experimental error. An open question
remains, however, whether the symmetry of the crystal is affected
by the magnetostriction. DTN has previously attracted interest
because the field-induced phase transition at $H_{c1}$ likely
belongs to the universality class of Bose-Einstein Condensation
(BEC). \cite{Zapf06} The tetragonal symmetry of the lattice
creates a necessary condition for conservation of the equilibrium
number of bosons, and therefore structural deviations away from
tetragonal crystal symmetry could disallow the Bose-Einstein
Condensation picture. Since the magnetostriction effects occur
gradually at fields above $H_{c1}$, the BEC picture would hold
right at $H_{c1}$ as reported, \cite{Zapf06} but become less valid
at high fields as the structure becomes increasingly distorted.
This possibility is currently being further investigated via
elastic neutron diffraction and ESR measurements.

It has been suggested that sound attenuation studies, which probe
magnon-phonon coupling, are another means of probing the magnitude
of $J(r)$. \cite{Cottam74} However, as demonstrated in
measurements of a similar antiferromagnetic quantum magnet
TlCuCl$_3$, \cite{Sherman03} the wave vector $k$ of the probing
phonons is vanishingly small compared to the wave vector of the
magnons, and thus the contribution of $J(r)$ to the magnon-phonon
coupling is negligible compared to the contribution of $D(r)$.

\begin{table}\caption{Tetragonal elastic moduli of DTN at room
temperature.}\label{elastic_constants}
\begin{tabular}{|c|c|}
  \hline
  \multicolumn{2}{|c|}{elastic moduli (GPa)} \\
  \hline
  $c_{11} = 26.1$ & $c_{12} = 15.3$\\
  $c_{33} = 14.2$ & $c_{44} = 11.2$ \\
  $c_{23} = 12.4$ & $c_{66} = 4.3$\\
  \hline
\end{tabular}
\end{table}

To our knowledge, the superexchange interaction in Ni-Cl-Cl-Ni
chains has not been previously investigated experimentally or
theoretically. For DTN, we speculate that the Cl-Cl bond
determines the magnitude of $J$ along the Ni-Cl-Cl-Ni chains,
since it is the weakest link, being nearly 2x longer than the
Ni-Cl bond (4.1 \AA{} vs 2.4 or 2.5 \AA{}). Early X-ray scattering
studies have also implied \cite{LopezCastro63} that the
lowest-energy lattice vibrations consist of the
NiCl$_2$-4SC(NH$_2$)$_2$ molecule moving as a unit, thus
supporting the idea that the Cl-Cl bonds that link adjacent
molecules are more susceptible to pressure than the Ni-Cl bonds
within a molecule.

 In summary, we have measured magnetostriction of
the organic quantum magnet NiCl$_2$-4SC(NH$_2$)$_2$ and we have
modelled the magnetostriction data by treating the compound as a
1-D magnetic system in which the strong dependence of the
superexchange interaction on the bond lengths along the c-axis
results in a magnetic stress. To our knowledge, this is the first
work in which the NN spin-spin correlation function is shown to be
directly proportional to an experimentally measurable quantity. It
also presents a new and straightforward method for determining the
spatial dependence of the exchange coupling over small distances.

\begin{acknowledgements}

This work was supported by the DOE, the NSF, and Florida State
University through the National High Magnetic Field Laboratory.
A.P.F. acknowledges support from CNPq (Conselho Nacional de
Desenvolvimento Científico e Tecnológico, Brazil). We would like
to thank S. Haas and N. Harrison for stimulating discussions.
\end{acknowledgements}


\end{document}